\documentstyle[aps,prl,epsfig]{revtex}

\def\beq{\begin{equation}}
\def\eeq{\end{equation}}

\begin{document}
\twocolumn[\hsize\textwidth\columnwidth\hsize\csname
@twocolumnfalse\endcsname
\title{Negative Magnetoresistance Produced by Hall Fluctuations in
a Ferromagnetic Domain Structure}
\author{Sergey\ V.\ Barabash and D. Stroud}
\address{Department of Physics,
The Ohio State University, Columbus, Ohio 43210}

\date{\today}

\maketitle

\begin{abstract}

We present a model for a negative magnetoresistance (MR) that would develop
in a material with many ferromagnetic domains even if the individual
domains have {\em no} magnetoresistance and even if there is no boundary
resistance.  The negative MR is due to a classical 
current-distortion effect arising
from spatial variations in the Hall conductivity, 
combined with a change in domain structure due to an applied magnetic field.
The negative MR can exceed 1000\% if the product of the
carrier relaxation time and the internal magnetic field due to spontaneous
magnetization is sufficiently large.

\end{abstract}

%\draft \pacs{PACS numbers: 75.80.+q, 75.30.Vn, 75.60.-d}
\vskip1.5pc]
 
\par
%\begin{twocolumn}

\newpage

%\section{Introduction}

There has recently been much interest in materials with a
{\em negative} magnetoresistance (MR).  In such materials, the resistivity in
a particular direction decreases when a magnetic field is applied.  For
example, the doped manganites show a very large decrease
known as colossal magnetoresistance\cite{helmolt,chahara,jin}.  
Similar behavior is achieved at comparatively small fields
in suitably strained manganite films\cite{wang,suzuki}.  While the
mechanism of these effects is not fully understood, it may arise from 
spin-dependent scattering at interfaces between domains, or within an
individual domain.

In this note, we discuss a model which could produce a
negative MR in a multidomain material even without any
spin-dependent or interfacial scattering.
This negative MR arises from what are sometimes called ``path length
effects'' (see, e.\ g., \cite{eckstein00}).
The model is based
on only two assumptions: (i) the medium is macroscopically inhomogeneous,
with a spatially varying Hall conductivity, and (ii) the inhomogeneous domain
structure depends on the applied magnetic field in a suitable way.  Given
(i) and (ii), one may extract the effective resistivity tensor $\rho_e$
of the medium as a function of field using standard methods as discussed, for 
example, in Ref.\ \cite{bergman}.  We will show that a simple, plausible model
for the conductivity tensors of the individual domains, combined with a suitable
method for calculating $\rho_e$, gives rise to 
a negative MR which can be very substantial.

We assume that there are two types of domains (two components), 
with magnetization parallel or antiparallel to the $z$ axis.
The two components have volume fractions of $p$ and $1-p$ respectively.
We carry out our calculation for two types of domain geometry\cite{note1}.  
In the first geometry (a ``random columnar'' microstructure), 
the domains have columnar symmetry with the columnar axis also lying along
the $z$ axis [see Fig.\ 1(a)]. 
This microstructure could represent a film 
with an out-of-plane easy magnetization axis
and a random distribution of domains.
The second geometry consists of a collection of parallel slabs arranged
%%CHANGE
perpendicular to the $x$ axis and infinite in the $y$ direction
[see Fig.\ 1(b)].
This geometry models domain structures obtained when the material is 
demagnetized in a certain way (see, e.\ g., \cite{helmolt}).

For both geometries, we assume that the 
ferromagnetic metal has a free electron conductivity
tensor with an effective magnetic field generated {\em internally}, by the
magnetization.  Thus, we assume that the internal field is $\pm B\hat{z}$
in the two components, where $\hat{z}$ is a unit vector in the direction normal
to the plane.  
The $3 \times 3$ conductivity matrices for the 
two components have elements
\begin{equation}
\sigma_{1,xx}=\sigma_{1,yy} = \sigma_{2,xx} = \sigma_{2,yy} = 
\sigma_0/[1+h^2]
\label{def:begin}
\end{equation}
\begin{equation}
\sigma_{1,xy} = -\sigma_{1,yx} = -\sigma_{2,xy} = \sigma_{2,yx} =
\sigma_0 h/[1+h^2],
\end{equation}
\begin{equation}
\sigma_{1,zz} = \sigma_{2,zz} = \sigma_0,
\end{equation}
with all other elements of $\sigma_1$ and $\sigma_2$ equal to zero.
Here $h = \omega_c\tau$ is a dimensionless magnetic field,
$\omega_c = qB/(m^*c)$ is the cyclotron frequency associated with 
the internal field B, 
and $\tau$ is a suitable relaxation time.

We will calculate the effective conductivity tensor in both geometries
using the standard effective medium approximation\cite{bergman}. (In the case
of the parallel slabs geometry, this approximation is, in fact,
exact\cite{bergman}.)  In both geometries, the self-consistency
condition takes the 
\newline

\begin{figure}
\centerline{
 \epsfig{file=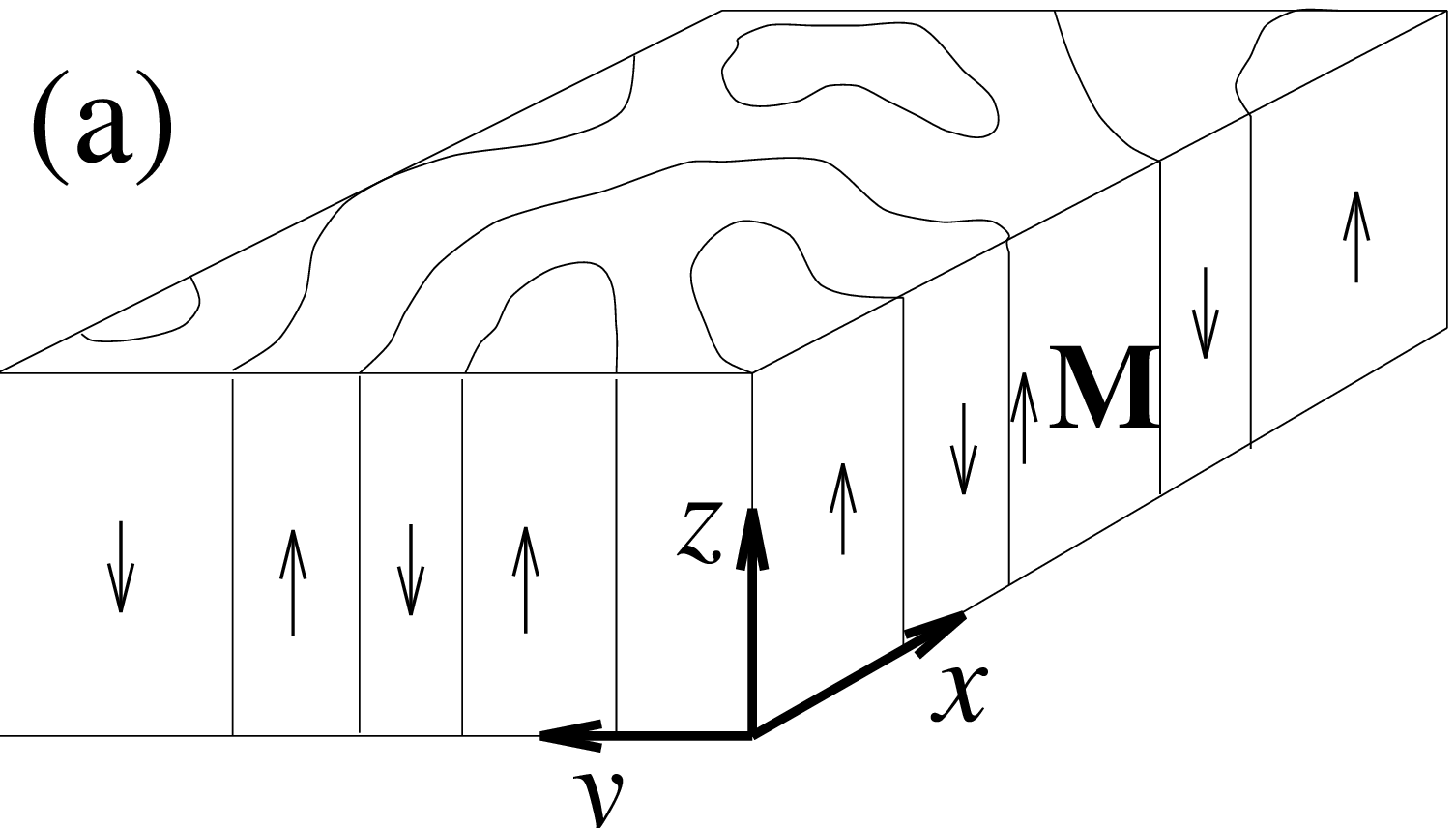, width=1.5in}
 \hskip0.5pc
 \epsfig{file=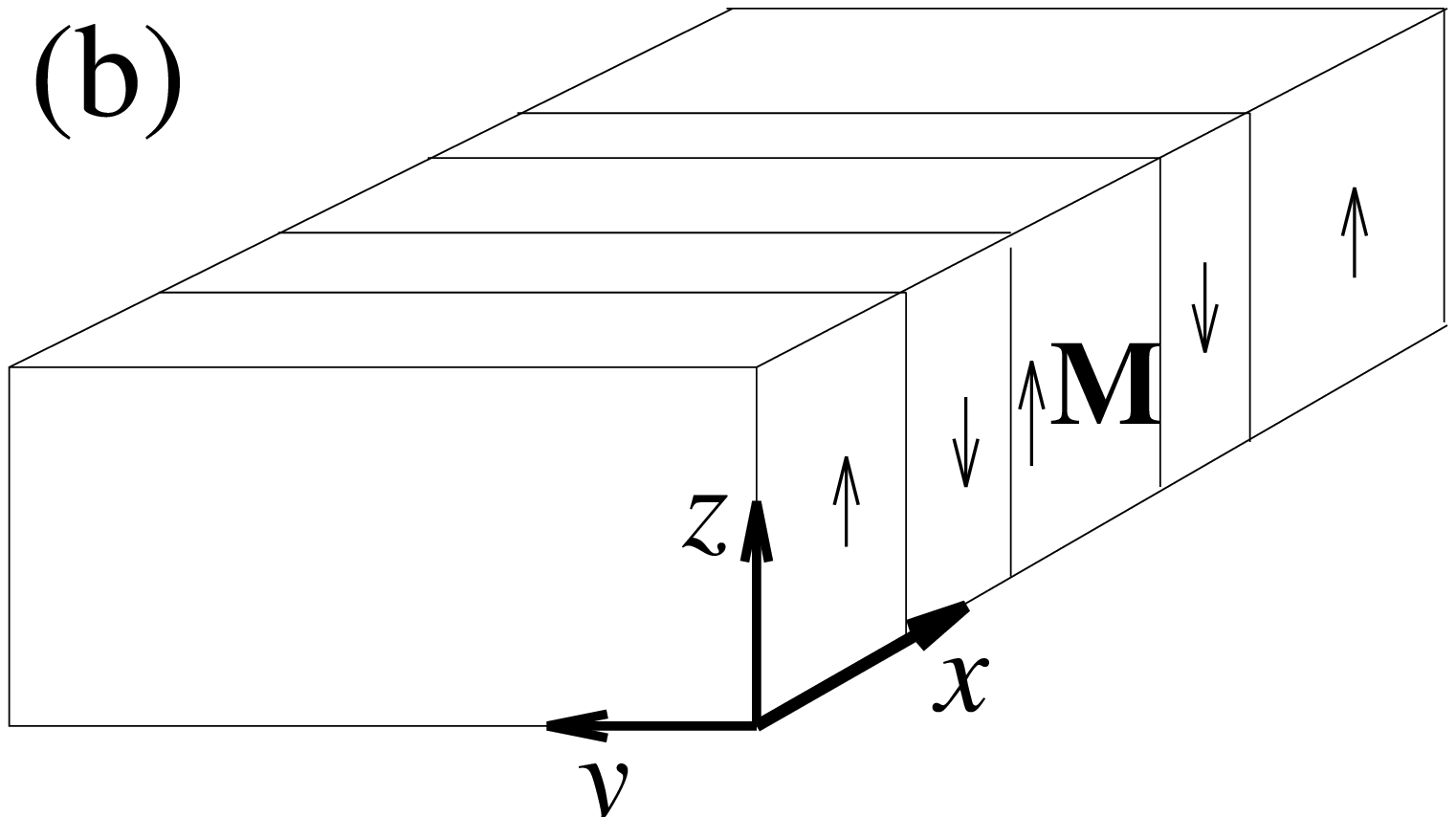, width=1.6in}
}
\vskip0.3pc
\caption{
Schematic of the discussed geometries.  (a) ``Random
columnar'' microstructure, consisting of
columnar domains with oppositely  directed magnetizations ${\bf M}$ 
and internal magnetic inductions ${\bf B}$ parallel to the column axes.
(b). ``Parallel slabs'' microstructure, with directions of ${\bf M}$ and internal
field ${\bf B}$ indicated by arrows.  The domains are separated by parallel
domain walls.  In both cases, ${\bf B}$ 
is assumed uniform within a domain and parallel to the domain walls, and,
in both cases, a thin film can be viewed as a slice of the microstructure
as shown.
}
\label{fig:stripes}
\end{figure}

\noindent
form
\begin{equation}
p\delta\sigma_1[1-\Gamma\delta\sigma_1]^{-1}
+(1-p)\delta\sigma_2[1-\Gamma\delta\sigma_2]^{-1} = 0.
\label{eq:ema}
\end{equation}
Here 
$\delta\sigma_i = \sigma_i - \sigma_e$, where $\sigma_i$ is the
conductivity tensor of the i$^{th}$ component and $\sigma_e$ is the effective
($3 \times 3$) conductivity tensor;  and $\Gamma$ is the 
depolarization tensor \cite{bergman}, which is different in the two geometries.

In the random columnar geometry $\Gamma$ is diagonal with diagonal
elements $\Gamma_{xx} = \Gamma_{yy} = -1/[2\sigma_{e,xx}]$; 
$\Gamma_{zz} = 0$, where we have used the fact that $\sigma_e$ is 
antisymmetric and that $\sigma_{e,xx} = \sigma_{e,yy}$.
The self-consistency condition (\ref{eq:ema}) can readily be multiplied out,
with the result
\begin{equation}
p\delta\sigma_1 + (1-p)\delta\sigma_2 = \delta\sigma_2\Gamma\delta\sigma_1.
\label{eq:ema1}
\end{equation}
This is a matrix equation for the effective conductivity tensor $\sigma_e$, whose 
components enter in (\ref{eq:ema1}) via both $\delta\sigma_i$ and $\Gamma$.
For our choice of conductivities, the solution to this equation is
\beq
\sigma_{e,xx}=\sigma_{e,yy}=
\frac{\sigma_{1,xx}} { \sqrt{1-4p(1-p)h^2/[1+h^2]}  },
\label{solution:xx}
\eeq
\beq
\sigma_{e,xy}=-\sigma_{e,yx}=
\frac{(2p-1)\sigma_{1,xy}} { \sqrt{1-4p(1-p)h^2/[1+h^2]}}.
\label{solution:xy}
\eeq
\beq
\sigma_{e,zz} = \sigma_{1,zz},
\label{solution:zz}
\eeq
with other components vanishing.

We first consider the special case $p = 1/2$.  Then (\ref{solution:xy})
reduces to $\sigma_{e,xy} = 0$, while (\ref{solution:xx}) is equivalent
to
\begin{equation}
\sigma_{e,xx} = \sqrt{\sigma_{1,xx}^2+\sigma_{1,xy}^2}.
\label{eq:sigexx}
\end{equation}
The in-plane resistivity is
$\rho_{e,xx} = [\sigma_e^{-1}]_{xx} = 1/\sigma_{e,xx}$,
where the last condition follows from the fact that $\sigma_{e,xy} =
-\sigma_{e,yx} = 0$.  Using our particular forms for $\sigma_{1,xx}$
and $\sigma_{1,xy}$, and using eq.\ (\ref{eq:sigexx}),
we finally get
\begin{equation}
\rho_{e,xx}\left(p = \frac{1}{2}\right) = \frac{\sqrt{1+h^2}}{\sigma_0}.
\end{equation}
For comparison, we can calculate the resistance for the case
$p = 1$ (or $p = 0$), which corresponds to a homogeneous magnetic 
material.
Since we have assumed a free-carrier conductivity tensor, the resistance
is simply
\begin{equation}
\rho_{e,xx}\left( p = 1\right) = \sigma_0^{-1}
\label{eq:rho(1)}
\end{equation}
Thus, if an applied field causes all the domains to line up parallel
with the field, the resistivity $\rho_{e,xx}$ will be reduced, i. e.,
there will be a {\em negative} MR.

Next, we consider the parallel slabs microgeometry.
In this case, the resistivity difference between the 
$p=1/2$ and $p=1$ cases may be even
larger.
For this geometry, the only nonzero element 
of the depolarization tensor
$\Gamma$ is $\Gamma_{xx} = -1/\sigma_{e,xx}$
\cite{note3}.
Carrying out the algebra in the self-consistency condition (\ref{eq:ema}),
we find that all the elements of 
$\rho_e \equiv (\sigma_e)^{-1}$ vanish except
\beq
\rho_{e,xx}= \sigma_0^{-1}\left[ 1+ h^2 -(2p-1)^2h^2\right]
\label{eq:slabxx}
\eeq
\beq
\rho_{e, xy}=-\rho_{e,yx}= - \frac {(2p-1)h} {\sigma_0},
\label{eq:slabxy}
\eeq
\beq
\rho_{e,yy}=\rho_{e,zz} = \frac{1}{\sigma_0}.
\label{eq:slabyy}
\eeq
In particular, at $p=1/2$ 
\beq
\rho_{e,xx}= \frac{ 1+ h^2} {\sigma_0},
\eeq
whereas at $p=1$ $\rho_{e,xx}$ is still given by (\ref{eq:rho(1)}).  
Note that in the columnar microgeometry $\rho_{e,xx}$ is the resistivity
perpendicular to the columns (or parallel to the film), while in the
parallel slabs microgeometry, $\rho_{e,xx}$ represents the
resistivity perpendicular to the slab surfaces.

Our picture of the negative MR in these structures
is now the following.  At zero (or very low) applied magnetic field, 
the sample  has either a random columnar or a slab 
microgeometry, with approximately equal numbers of up and down domains.
The resistivity then corresponds to $ p = 1/2$.
As the applied field is increased, the domains parallel to the field
grow at the expense of the antiparallel domains, so that eventually
the resistivity is close to that for $p = 1$.
The total fractional change in $\rho_{xx}$ (with respect to the 
low resistance value) is just
\begin{equation}
\frac{\Delta \rho}{\rho} 
 \equiv \frac{\rho( p = 1) - \rho(p = 1/2)}{\rho(p = 1)}
 = 1-\sqrt{1+h^2} < 0
\label{eq:EMAchange}
\end{equation}
for the random columnar domain structure; and
\beq
\frac{\Delta \rho}{\rho} =-{h^2}
\label{eq:slabchange}
\eeq
for the parallel slabs domain structure.  Both of these
correspond to a {\em decrease} in resistivity.  

%\section{Numerical Examples and Discussion}

To compare these analytic results to experiments, the quantity $2p-1$ 
must be related to the external magnetic field ${\bf H}$.   
Typically, in a sample with domains, the average sample 
magnetization ${\bf M_e}$ is a hysteretic function of ${\bf H}$.  
As an illustration, we model this function by
$ M_e= M_{sat} \tanh\left( (H\pm H_{cr})/H_{sat}\right)$, where $H_{cr}$ is
the coercive field, $H_{sat}$ and $M_{sat}$ are characteristic saturation
values of the external field and magnetization, and the sign choice 
depends on the direction in which the hysteresis loop is traversed.
If the {\em local} magnetization can have only the two values $\pm M_{sat}$ in
the up and down domains, then
$M_e=M_{sat}(2p-1)$, and hence
\beq
2p-1= \tanh \left( \frac{ H\pm H_{cr} }{H_{sat}} \right).
\label{eq:hyster}
\eeq
\begin{figure}
 \centerline{
  \epsfig{file=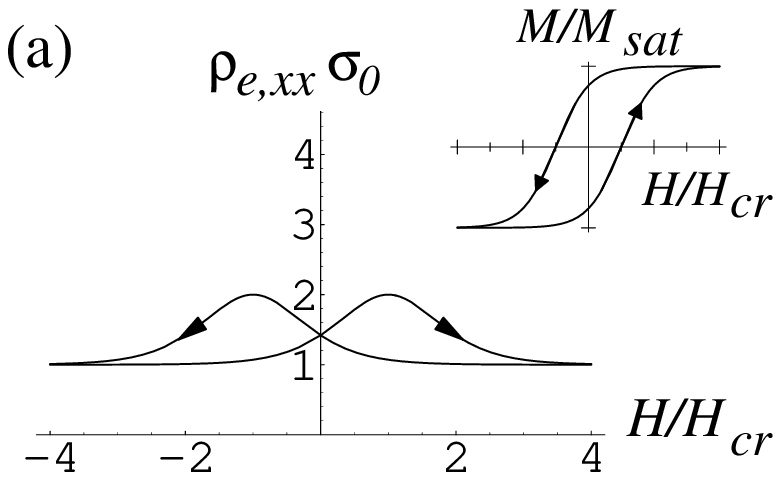, width=1.7in}
  \epsfig{file=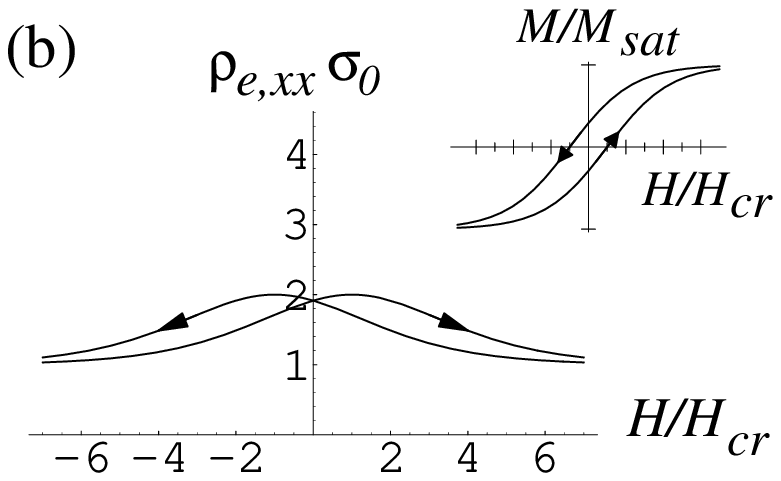, width=1.7in}
 }
\vskip0.5pc
 \centerline{
  \epsfig{file=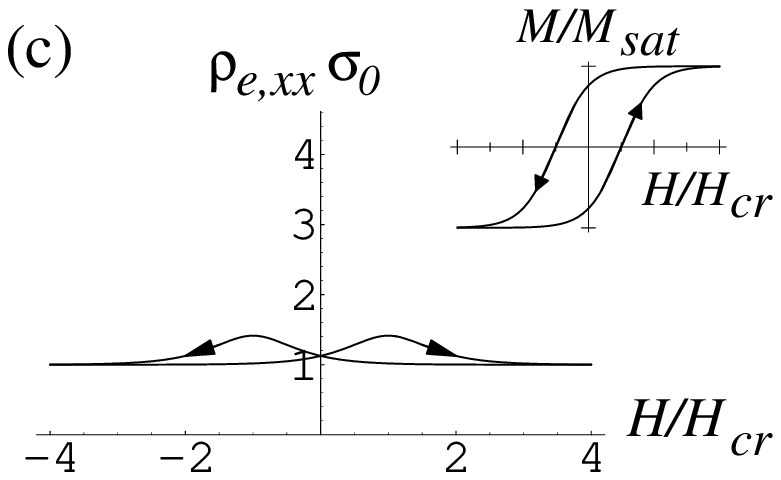, width=1.7in}
  \epsfig{file=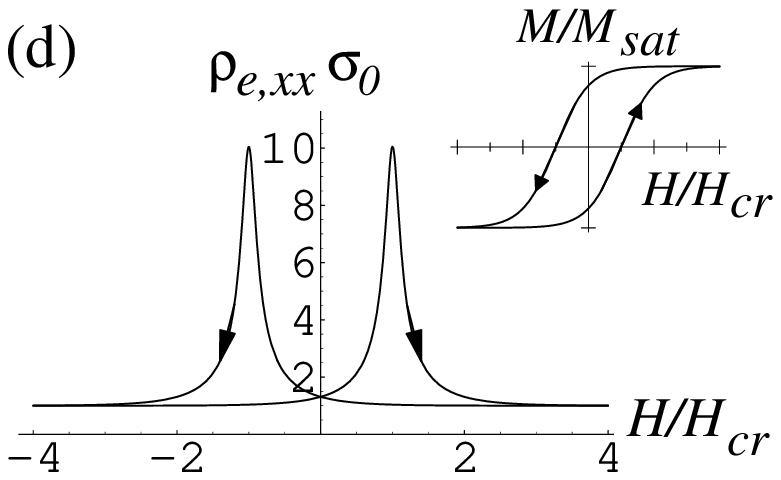, width=1.7in}
 }
 \caption{
Predicted behavior of the effective resistivity $\rho_{e,xx}$,
plotted versus applied magnetic field ${\bf H}$.
(a) and (b): parallel slab microgeometry, [eqs.\ 
(\ref{solution:xx}-\ref{solution:xy})]; 
(c) and (d): random columnar microgeometry
[eqs.\ (\ref{eq:slabxx}-\ref{eq:slabyy})].  The plots are shown for
several different values of the parameters $H_{cr}/H_{sat}$
and $\omega_c \tau$, using eq.\ (\ref{eq:hyster}): (a) and (c) $H_{cr}/H_{sat}=1$, $\omega_c \tau=1$;
(b) $H_{cr}/H_{sat}=0.3$, $\omega_c \tau=1$;
(d) $H_{cr}/H_{sat}=1$, $\omega_c \tau=10$.  Insets show 
corresponding magnetization hysteresis loops, also obtained using eq.\
(\ref{eq:hyster}).
}
 \label{fig:results}
\end{figure}

\noindent
We have used this expression in the above formulas for $\rho_{e,xx}$
to obtain the hysteresis in $\rho_{xx}(H)$;
the results are shown in Fig.\ \ref{fig:results}.  Evidently 
both the shape
and the positions of the peaks in $\rho_{e,xx}(H)$
depend on the domain microgeometry (i. e., whether it is ``random columnar''
or ``parallel slabs'') and the squareness of the hysteresis loop (controlled
by the parameter $H_{cr}/H_{sat}$).

The {\em magnitude} of  
$ \Delta \rho_{xx}/\rho_{xx}$
is entirely controlled by the parameter
$h=\omega_c\tau$ [cf. eqs.\ (\ref{eq:EMAchange})
and (\ref{eq:slabchange})].  
At room temperature\cite{roomTnegativeMR}, $h$ is likely to be small: assuming
$\tau = 2 \times 10^{-14}$ s at $T = 300K$ 
(1/10 that of Cu), a local magnetization $M \sim 10^3$Gauss and 
internal $B \sim 4 \pi M$, we find 
$\omega_c \tau \sim 10^{-3}$.  But at low temperatures,
$h$ could be very large: again assuming $\tau \sim 0.1\tau_{Cu}$ at
$T = 4K$, we find $\omega_c \tau \sim 10^2$, implying a very large negative 
$\Delta \rho_{xx}$, especially in the slab geometry. 
In reality both $B$ and $\tau$ are likely to be reduced
near the domain walls\cite{note2}; nevertheless,
the  change in $\rho_{xx}$ could easily exceed 1000\%, according to eqs. 
(\ref{eq:EMAchange}) and (\ref{eq:slabchange}).

Remarkably, these results resemble experimental observations in 
strained manganite films \cite{wang}. In these films, the
magnetoresistance is large and negative
at low $T$, but is negligible above 100K.  
Moreover, our
prediction (see Fig.\ \ref{fig:results}), that the magnetic
field of the peak resistance is
close to the coercive field, is also consistent
with experiment.  
Of course,there are certainly
other, probably more 
significant, mechanisms contributing to the negative MR
in manganite films (such as colossal magnetoresistance or spin-dependent 
scattering).  Nonetheless, it seems plausible that this Hall-induced
negative MR may play a significant role in some materials.

In conclusion, we have demonstrated that in a ferromagnet with
a suitable domain structure, the Hall effect arising from the local
magnetic field will increase the effective resistivity of the material.
When an external magnetic field destroys the domain structure, it will
also reduce the resistivity, thereby generating a substantial negative
MR.  The magnitude of this effect is controlled by the
parameter $\omega_c\tau$, which may be very large at low temperatures,
and by details of the domain microgeometry, as well as by the shape of
the hysteresis curve $M(H)$.

This work has been supported by NSF Grant DMR97-31511.  We thank 
S.\ A.\ Batra for useful conversations.


\begin{thebibliography}{99}
\bibitem{note1}
We assume that the magnetization is normal to the film surface.
In the opposite case of in-plane magnetization, the
results for the slab geometry would still be exact, 
provided that the domain diameter $d$ is smaller than the film
thickness $t$. 
The latter condition will probably not be satisfied in very thin films,
as one expects $d \sim \sqrt t$\cite{Landau8}.
When $d \gg t$ and the magnetization is in the plane of the film
(as, e.\ g., in ref.\cite{mathur}),
the effective film resistivity $\rho_{e,xx}$ will depend 
on the number and positions
of the domain walls, rather than on the volume fractions $p$ and $1-p$
of the domains as it does in our model.
\bibitem{note3}
For scalar conductivities, these values would lead to the familiar
result that resistivities add in the direction perpendicular to the
slab interfaces, while conductivities add in the parallel directions.
\end{thebibliography}
\end{document}